\documentclass{aastex} 		
\usepackage{emulateapj5}	
\usepackage{apjfonts}		
\usepackage{natbib}
\usepackage{graphicx}
\bibpunct{(}{)}{;}{a}{}{,}

\newcommand{\lina}{\ion{Fe}{1}~$\lambda$6302.5~\AA}
\newcommand{\linb}{\ion{Fe}{1}~$\lambda$6301.5~\AA}
\newcommand{\linc}{\ion{Fe}{1}~$\lambda$15648~\AA}
\newcommand{\lind}{\ion{Fe}{1}~$\lambda$15652~\AA}
	\newcommand{\modified}[1]{{#1}}

\shorttitle{Simultaneous visible and IR observations of solar internetwork regions}
\shortauthors{S\'anchez Almeida et al.}

\begin{document}
   \title{Simultaneous visible and infrared spectro-polarimetry\\ 
	of a solar internetwork region
	}

    \author{J. S\'anchez Almeida$^1$,
    	I. Dom\'\i nguez Cerde\~na$^2$,
	{\sc and}
    	F. Kneer$^2$} 
    \affil{$^1$Instituto de Astrof\'\i sica de Canarias, 
              E-38205 La Laguna, Tenerife, Spain}
   \affil{$^2$Universit\"ats-Sternwarte,
              Geismarlandstra\ss e 11, D-37083 G\"ottingen, Germany}
   \email{jos@ll.iac.es, ita@uni-sw.gwdg.de, kneer@uni-sw.gwdg.de}

\begin{abstract}
We present the first
simultaneous infrared (IR) and visible
spectro-polarimetric observations
of a solar internetwork region.
The Fe~{\sc i} lines at
$\lambda$6301.5~\AA,
$\lambda$6302.5~\AA,
$\lambda$15648~\AA,
and $\lambda$15652~\AA\
were observed,
with a lag of only 1 min, using
highly sensitive
spectro-polarimeters operated in two different telescopes (VTT
and THEMIS at the {\em Observatorio del Teide}).
Some 30\% of the observed region shows IR and
visible Stokes $V$
signals above noise.
These polarization signals indicate the presence of kG
magnetic field strengths (traced by the visible lines) co-existing with
sub-kG fields (traced by the infrared lines).
\modified{
In addition, one quarter of the pixels with signal
have visible and IR Stokes $V$ profiles with opposite polarity.
}
We estimate the probability density function
of finding each
longitudinal
magnetic field strength in the region.
It has a tail of kG field strengths that accounts for most of the (unsigned)
magnetic flux of the region.
\end{abstract}
\keywords{
          Sun: magnetic fields --
          Sun: photosphere}

%
%

\section{Introduction\label{introduction}}

Most of the solar surface appears non-magnetic when it is observed
in routine synoptic magnetograms (e.g., those
obtained at Kitt Peak\footnote{http://www.noao.edu/kpno/}
or with the MDI instrument on board of
SOHO\footnote{http://sohowww.nascom.nasa.gov/}). However,
magnetic fields are detected almost everywhere,
also in internetwork (IN) regions,
when the polarimetric sensitivity and the angular
resolution  exceed a threshold.
These {\em magnetic fields
of the quiet Sun}, elusive but now accessible to many
existing spectro-polarimeters,
have received an increasing interest
during the last years. They produce only weak signals,
but they cover much of the solar surface and, therefore,
they may be carrying most
of the unsigned magnetic flux and energy existing
on the solar surface at any given time
\citep[e.g.,][ and references therein]{san02b}.
This  makes the IN potentially important 
to understand
the global magnetic properties of the Sun.

The observational study of the IN fields
began in the seventies \citep{liv75,smi75}, but it
is still in an initial phase
due to the complexity of the magnetic topology.
Opposite polarities seem to  coexist
in one and the same resolution element of typically 1\arcsec 
\citep[see][]{san96,san00,sig99,lit02,kho02}. 
In addition, large systematic differences 
of intrinsic field strength are deduced
depending on the diagnostic technique.
Techniques based on visible lines 
favor kG field strengths 
\citep{gro96,sig99,san00,soc02,dom03a,dom03b}, whereas
inferences based on infrared (IR) lines indicate
sub-kG field strengths \citep{lin99,kho02}. (We add for completeness that 
estimates based on the Hanle
effect indicate the presence of even weaker fields, e.g.,
\citealt{fau95,bia99}.)
\citet{san00} conjectured that this systematic difference
actually
reflects the existence of a continuous
distribution of field strengths ranging from sub-kG to kG
\citep[see also][]{soc02}. Different observations
are biased towards a particular part of such a distribution,
producing the observed systematic difference of
field strengths. Even more, \citet{san00}
put forward a specific physical mechanism
that biases the IR observations towards sub-kG fields.
It is caused by the vertical gradient of field strength
existing in any magnetic structure (provided they
satisfy some sort of mechanical balance with the mean
solar photosphere).
The Zeeman splitting of the IR lines is
so large that the absorption in various layers, with different field strengths,
occurs at substantially different wavelengths. The absorption is thus
spread over a wide range
of wavelengths, which dilutes the polarization signals.
On the other hand, sub-kG fields do not disperse the IR absorption.
In this case the IR lines are in the so-called weak field regime
so, independently of the field strength, the absorption is always
produced at the same wavelengths, and
a significant polarization signal builds up.
\citet{soc03} show that the existence of a
{\em horizontal} gradient 
\modified{
can also explain  
}
the observed discrepancy.
In this particular case  a distribution of field
strengths is chosen such that
the visible lines reflect kG fields
because most of the magnetic
flux is in the form of kG fields.
In the IR lines, however,
the kG signal is much reduced because of the wavelength
smearing effect described above.

 So far the seemingly contradictory visible and IR observations were
taken independently (by different authors at different times).
If the conjecture on the coexistence
of weak and strong fields were correct, a simultaneous
observation of the visible and IR lines would have
to give co-spatial signals showing both, signatures of
kG (visible) and of sub-kG (IR) fields. The
present study was
meant to test this specific prediction.

The simultaneous observations and few
details of the data reduction
are described in \S~\ref{observations}.
Then we explain
the data analysis, the classification of the polarimetric
profiles, and the determination
of magnetic field strengths
(\S~\ref{data}).
We present the main results in \S~\ref{result}, and the
conclusions in \S~\ref{conclusions}.

\section{Observations\label{observations}}

We took
advantage of the Spanish Observatorio del Teide (Tenerife, Spain) having
two leading telescopes
with sensitive
spectro-polarimeters. The IR observations were carried out with the Tenerife
Infrared Polarimeter \citep[TIP,][]{mar99}
at the German Vacuum Tower
Telescope (VTT). We obtained spectra of all four Stokes parameters for
the IR lines
\linc~(effective Land\'e factor, $g_{eff}=3$)
and \lind~($g_{eff}=1.53$).
The observations in the visible range were gathered with the
spectro-polarimetric mode of the THEMIS telescope
\citep{lop00}.
We chose 3 different
spectral ranges for the visible observations.
Only one
containing the lines \linb~($g_{eff}=1.67$) and
\lina~($g_{eff}=2.5$)
is analyzed here.
The slit width was set to 0\farcs5  in both telescopes,
and we scanned the solar surface
from east to west with
a step size of 0\farcs5.
Each scan consisted of 60 positions, with an integration
time of 30~sec per position.

\begin{figure}
\epsscale{0.5}
\plotone{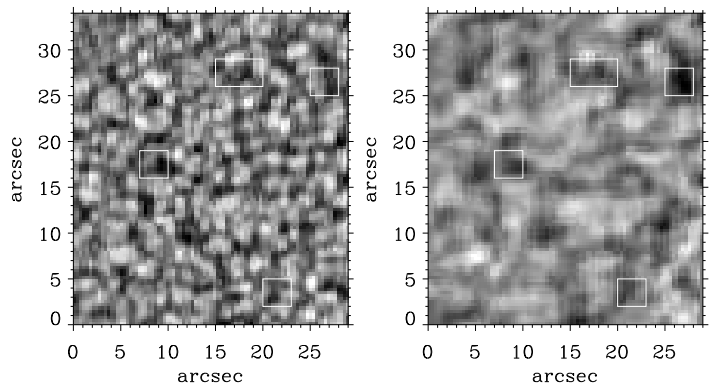}
\caption{Continuum intensity images from the IR (left) and the
  visible spectro-polarimetric scans (right). The
  squares point out structures which can easily be
  identified in both images demonstrating the good co-alignment.
  We scanned along the horizontal direction (solar East-West); the vertical
  direction is the direction along the spectrograph 
  slit (North-South).
  }
\label{img}
\end{figure}

The dataset for this work was obtained on August 10, 2002,
in a quiet Sun region close to disk center
($\mu=0.94$).
We selected IN regions trying to avoid
bright features in the  H${\alpha}$ and
\ion{Ca}{2}~K
slit-jaw filtergrams supplied by the VTT.
Having a few network patches in our fairly large Field-of-View (FOV)
is unavoidable, though.
In order to
point both telescopes to the
same region, we
benefited from
a video link
that provided VTT slit-jaw images
at THEMIS.
First, we pointed
both telescopes to an easily identifiable structure,  e.g.,
a sunspot.
Then we moved the
VTT pointing to a quiet IN region and displaced THEMIS
by the same distance.
For the data set analyzed here, 
this method of {\em differential pointing}
gave an offset of the co-alignment of
the two telescopes of 1\arcsec. 
\modified{
When the offset is corrected
it leads 
}
to a time lag between the IR and visible data of only 1 
minute.
We found this offset by
shifting the
continuum intensity images obtained from the
visible and the IR scans (Fig. \ref{img}).
The optimum shift is chosen as that
yielding the minimum difference
between the two images.\footnote{Actually,
we allow for
a shift, a rotation
of the FOVs,
and a scale factor in the direction
along the
slit. The rotation is only 1\degr, whereas the scale
factor is consistent with the ratio of effective focal lengths
in the focal plane of the two polarimeters.}
Figure \ref{img} shows
the resulting
intensities in the IR
(left) and the
visible (right). 
Despite the difference in angular resolution,
it is clear
that both images correspond to the same region. The white
boxes point out a set locations on the Sun,
and one can easily identify the same structures in the boxes
of the two spectral ranges.
The spatial resolution for the IR observations is 1\arcsec--1\farcs2, with a
granulation contrast of 1.7\%, and
it becomes 1\farcs5--1\farcs7
in the visible, with a
contrast of  1.9\%.
The resolution was estimated
as the cutoff of
the power spectra
of the continuum images shown in Fig. \ref{img}.
The IR
resolution is better due to the use of
an active tip-tilt mirror
that removes image motion
during scanning
\citep[][]{bal96}.
The FOV common to visible and IR observations
is of the other of 
30\arcsec~$\times$~35\arcsec\ (Fig.~\ref{img}).

The data reduction for the IR and visible ranges was very similar:
flat field correction,
removal of  fringes,
demodulation to get the Stokes parameters from the measurements,
and
correction of
seeing-induced crosstalk
by combining the two beams of each polarimeter.
The instrumental polarization (IP) of the VTT was
cleared away
using a model Mueller matrix for the telescope,
which was calibrated with large linear polarizers inserted at the
telescope entrance
\modified{
\citep[][]{col03}.
}
THEMIS data do not need this step since it is an
IP free telescope.
The reduction process renders Stokes $I$, $Q$, $U$ and $V$ profiles
in both spectral ranges for each
pixel
of our FOV.
Here we only  use
intensity and  circular polarization.
The noise in the polarization
measurements was estimated as $3-4\times 10^{-4}~I_{\mathrm c}$
in the IR data, and $7\times 10^{-4} I_{\mathrm c}$ in
the visible data ($I_{\mathrm c}$ is the continuum intensity).
They correspond to the rms fluctuations of the
polarization in continuum wavelengths.

\section{Data analysis\label{data}}

We select for analysis those IR and visible Stokes $V$ profiles
whose peak polarization
is larger than three times the noise level.
This thresholding separates
clear signals from very noisy profiles or pure noise. The pixels with
both IR and visible
signals above 
their respective
threshold
correspond to 30\% of the FOV. However,
an additional 40\% of the pixels
present polarization signals
only in either the visible or the IR. This study
deals with
those pixels with signals in both spectral ranges 
\modified{
(some 1840).
}

All selected profiles were classified using a Principal Component
Analysis (PCA)
algorithm,
similar to the one used by \citet[ \S~3.2]{san00}. We classify,
simultaneously, the IR and visible Stokes $V$ profiles.
All vectors
are scaled
to the maximum of the blue lobe of the \lina\ Stokes~$V$ profile.
The PCA classification results in 25
different classes
with more
than 0.5\%
of the 
\modified{
classified 
}
pixels,
and 15 classes containing rare
profiles (between 0.5\% and 0.2\%).
Figure \ref{clas} shows some of the significant classes. We note, firstly, that
the IR profiles in class 0 are similar to the
most abundant class
obtained by
\citet[][ Fig.~1]{kho02}.
Secondly, class 2 corresponds
to network structures in our FOV, and it has characteristic broad IR
profiles
(also observed in plage regions by \citealt{ste87}).
Finally,
the most conspicuous fact is that
some classes (e.g., \# 4 and \#~8 in Fig.~\ref{clas})
show {\em opposite} polarity in the visible and in the IR lines
(the lobes have opposite sign).

In order to estimate
magnetic field strengths, we performed  
Milne-Eddington (ME) inversions of the profiles
of all individual  pixels, as well as
the mean profiles among those in each
class.
We use MILK,
a code which is part of the 
Community Inversion Codes\footnote{http://www.hao.ucar.edu/public/research/cic/}
developed by HAO \citep[see][]{soc01}.
Each inversion provides 11 parameters, including the
longitudinal component of the
magnetic field
strength, and the fill factor (fraction of resolution
element covered by magnetic fields).
(For a description of the procedure and approximations, see, e.g.,
\citealt{sku87}.)
The inversion is
separately done for the two
pairs of lines, i.e.,
we carry out two independent inversions per pixel,
one for the visible pair and another for the IR
pair. In each inversion we use the same atmospheric and magnetic field
parameters for
the two lines, whereas
the ratio of absorption coefficients is fixed
according to the expected values.
Note that our ME inversion does not fit line asymmetries,
although they are clearly
present among the observed profiles.

\section{Results\label{result}}

The PCA classification indicates
that 25\% of the pixels 
\modified{
with coincident visible and IR signals
}
give
visible and IR profiles with opposite polarity.
It follows that the
visible and IR lines are tracing 
different magnetic structures in the same
resolution element.
Some
classes have
IR profiles with three lobes while the visible profiles  have
only two
(not included in Fig.~\ref{clas}).
Actually, the observed Stokes $V$ profiles present
a wide variety of asymmetries,
in agreement with previous observations
\citep[e.g.,][]{san00,kho02}.

The ME inversions show that most of the classes have visible Stokes $V$
profiles characteristic of kG fields,
while, simultaneously, the IR profiles
indicate the presence of
sub-kG field strengths.
This also happens when we analyze all individual profiles. The
result
cannot be significantly biased by noise,
since the mean profile of
a class results from averaging hundreds of
profiles and, therefore, bears negligible noise (see Fig. \ref{clas}).
On the other hand, the systematic difference between the
IR and the visible
is insensitive to the specific technique used to diagnose
the field strength.
The ratio between the
Stokes $V$ extrema of the two visible lines
can be used to diagnose
the regime of 
field strength in the atmosphere. A ratio close to 1 indicates kG magnetic
fields,
whereas a value of some
0.6 is expected for sub-kG fields
\citep[see Sect. 5 in][]{soc02}. Most of the classes have ratios
between 0.9 and 1,
which evidences IN
magnetic structures in the regime of kG fields.
After the inversion of profiles in the various classes,
and also from individual profiles, we
found a mean magnetic field strength of 1100 G from the visible lines.
They fill only 1\% of the resolution elements.
The mean field strength in the IR
is $300$~G,
and covers some 2\% of the
part of FOV studied here.

The ME inversions also show a few classes with weak
magnetic fields both in
visible and the IR.
There are also classes
with strong fields in both spectral ranges
corresponding to network points in our FOV (class 2 in Fig. \ref{clas}).

Longitudinal magnetograms measure
flux density.
One can  infer it from ME inversions as the  product of the
longitudinal component of the field strength
times the fill factor. When averaged over the 30\%
of the FOV that we study, the mean unsigned flux density
turns out to be  $11$~G in the visible, and $6$~G
in the IR.
	Due to the complexity of the IN magnetic fields,
	the observed unsigned flux is expected to increase
	with increasing
	angular resolution \citep[see, e.g.,][]{san03,dom03b}.
	The fact that the IR, with higher resolution, has
	lower unsigned flux, seems to indicate that this is
	a true solar effect. Specifically, most of the
	unsigned  magnetic flux of the IN seems to be in the form
	of kG fields.
On the
other hand,
the {\it signed} flux
of the region
in the visible is
$-1.5$~G, while
it is less than $0.2$~G
in the IR.
A similar
{\em net}
flux has been observed
before in visible lines
\citep[e.g.][]{lit02,dom03b}.
We find it to
be co-spatial with
very low  signed flux in IR lines.

Using magnetic fields from the ME inversion, we have estimated the
Probability  Density  Function (PDF)
of finding a given longitudinal magnetic field
strength in our IN quiet Sun region.
We assume that the visible and the
IR polarization signals trace different structures
(a reasonable hypothesis, in view of the findings
described above). The range of observed
field strengths is divided in bins of 50~G. Then,
treating visible
and IR observations as independent measurements,
we assigned each measured longitudinal magnetic field
to one of the bins.
The probability of finding 
\modified{
magnetic field strengths
}
in a given bin (i.e., the PDF) will  be
proportional to the sum of fill factors
of all those  measurements
in the bin. The result is shown in Fig. \ref{pdf},
the solid line.
This PDF has been normalized so that
its integral is 0.3, i.e., the fraction of
FOV under analysis. In other words, the normalization
assumes that the remaining 70\% of the surface is
field free.
Figure \ref{pdf}
includes the PDF
for the IR data alone
(the dashed line), and for
the visible data alone
(the dotted line).
We also plot the PDF suggested by \citet[ the dash-dotted line]{soc03}.
We find that 75\% of the total flux is concentrated in fields
stronger than $500$~G, which, however, occupy only
0.4\% of the full FOV.

\modified{
A final comment is in order. The visible and IR maps
that we employ have a time lag of one minute and
were coaligned with a finite accuracy (however better than the 
angular resolution). The lack of strict co-alignment and simultaneity 
may produce some differences between the visible and the IR spectra, 
however, they would be of random nature  and, consequently,
they can hardly explain the observed systematic differences. 
}

\section{Conclusions\label{conclusions}}

The (almost) simultaneous visible and IR
Stokes $V$ spectra
described above have several
conspicuous properties.
They  reveal the coexistence of kG and sub-kG magnetic
fields in most
of the pixels
with polarization signals above the noise (30\% of the FOV).
%
%
The IR lines
indicate
sub-kG fields whereas visible lines trace  kG.
This demonstrates the
conjecture by
\citet{san00} and \citet{soc02,soc03}.

\modified{
At least 25\% of pixels with coincident visible and IR signals
have one polarity in the visible  and the opposite in the IR.
}
This second result offers a direct proof of the
existence of two polarities in 
\modified{
1\farcs 5\ 
}
resolution elements. They have been inferred from line asymmetries \citep{san00},
as well as from the skewness of the distribution of line-of-sight
magnetic field inclinations
in measurements
away of
the solar disk center \citep{lit02}.

\begin{figure}
\epsscale{0.5}
\plotone{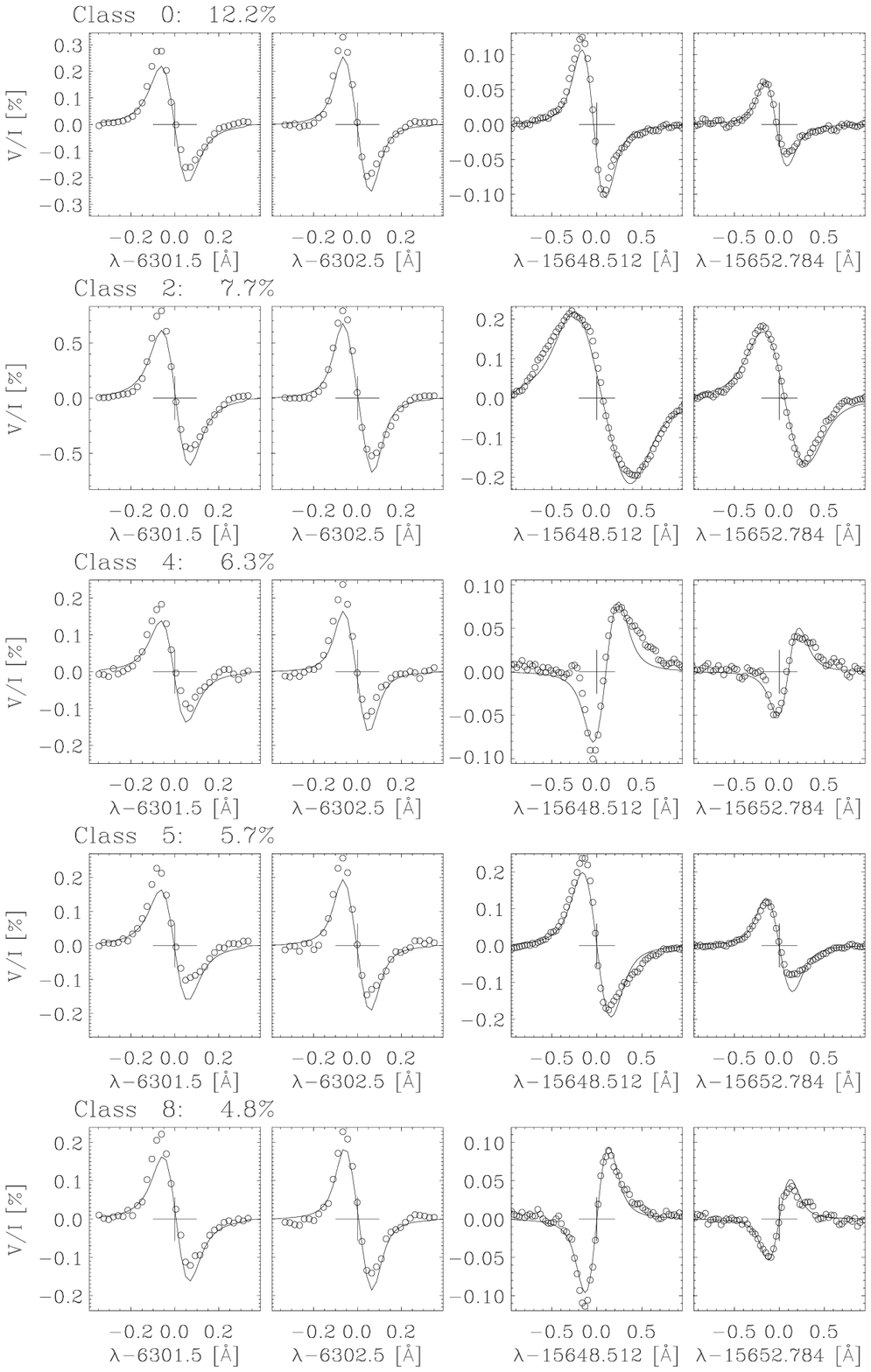}
\caption{Representative Stokes $V$ profiles from the
	PCA classification.
  	Mean profiles of \linb~(left), \lina , \linc , and
  	\lind~(right) among those in classes 0, 2, 4, 5 and 8
	(see the labels on top of the figures, where the percentage
  	of profiles included in each class is also
	given).
	Wavelengths are in \AA\ relative to the
	line center (indicated in each figure as a large plus sign).
	The small circles represent actual observations whereas the
	solid lines correspond to ME fits to the observed data.
	}
\label{clas}
\end{figure}

\begin{figure}
\epsscale{0.4}
\plotone{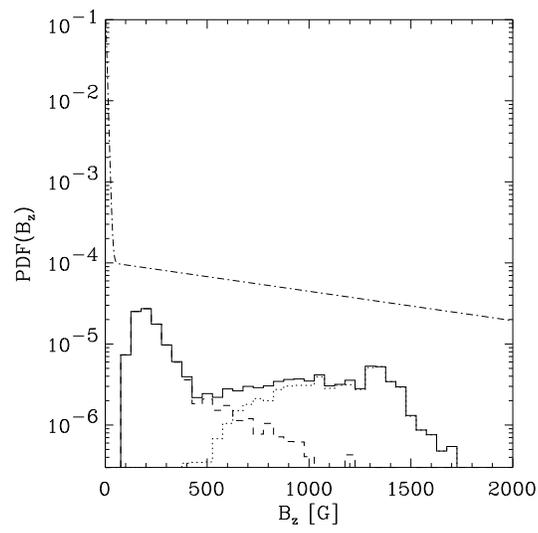}
\caption{Probability Density Function (PDF)
	for observing a longitudinal magnetic field strength
	$B_{\mathrm z}$.
	The solid line includes both IR and visible data, whereas
	the distribution of values obtained solely from the IR lines
	and the visible lines are shown as the dashed and the
	dotted lines, respectively.
The dot-dashed PDF was suggested
by \citet{soc03}
and it is represented for reference.
}
\label{pdf}
\end{figure}

The Probability Density Function for the
longitudinal magnetic fields, that we infer from
ME inversions, has a long tail of kG fields.
Some 75\% of the observed (unsigned) magnetic flux is in
field strength larger than 500~G. This excess
of kG flux is in agreement with the PDF
suggested by
\citet{soc03}
to reconcile discrepancies
between field strengths obtained
with different methods (see \S~\ref{introduction}).
We have also found the extended tails in the IR line \linc\ predicted by
such PDF
(see classes 0 and 5 in Fig. \ref{clas}). They carry
the information on the strong
fields, but are
easily misinterpreted due to the large
lobes formed by weak fields.

Our observations 
\modified{
prove
}
that the use of
only visible lines or only IR lines
to study the IN fields
biases the results.
Diagnostics based on
simultaneous visible and IR observations
offer a viable and reliable alternative.

\acknowledgements
We thank M. Collados for training with TIP operation,
and for assistance during the data reduction.
Thanks are also due to H. Socas-Navarro, who provided
the IC,
to T. Eibe, who supported us with TIP at the
VTT,
and to the team of
support astronomers and operators of THEMIS.
\modified{
Comments on the manuscript by E. Landi Degl'Innocenti and the 
referee were very useful.
}
IDC acknowledges support by the Deutsche Forschungsgemeinschaft (DFG) through
grant 418\,SPA-112/14/01.
The VTT is operated by the Kiepenheuer-Institut f\"ur
Sonnenphysik, Freiburg, and the French-Italian telescope THEMIS is operated
by CNRS-CNR, both at the Observatorio del Teide of the
Instituto de Astrof\'\i sica de Canarias, Spain.
The work has partly been funded by the Spanish Ministry of Science
and Technology, 
project AYA2001-1649.
%
%

%

\end{document}